\documentclass[aps, twocolumn, a4paper, superscriptaddress, 10pti, pre]{revtex4-2}

\usepackage{graphicx}
\usepackage{amsmath, amssymb}
\usepackage{lmodern}
\usepackage[T1]{fontenc}
\usepackage[utf8]{inputenc}
\usepackage[english]{babel}
\usepackage{xcolor}

\begin{document}

\title{Chaotic behavior in Lotka-Volterra and May-Leonard models of biodiversity}
\author{D. Bazeia}
\affiliation{Departamento de Física, Universidade Federal da Paraíba, 58051-970 João Pessoa, PB, Brazil}
\author{M. Bongestab}
\affiliation{Departamento de Física, Universidade Federal da Paraíba, 58051-970 João Pessoa, PB, Brazil}
\author{B. F. de Oliveira}
\email{bfoliveira@uem.br}
\affiliation{Departamento de Física, Universidade Estadual de Maringá, 87020-900 Maringá, PR, Brazil}

\begin{abstract}
Quantification of chaos is a challenging issue in complex dynamical systems. In this paper, we discuss the chaotic properties of generalized Lotka-Volterra and May-Leonard models of biodiversity, via the Hamming distance density. We identified chaotic behavior for different scenarios via the specific features of the Hamming distance and the method of \textit{q}-exponential fitting. We also investigated the spatial autocorrelation length to find the corresponding characteristic length in terms of the number of species in each system. In particular, the results concerning the characteristic length are in good accordance with the study of the chaotic behavior implemented in this work.		
\end{abstract}

\maketitle

{\bf The use of lattice based stochastic models, such as May-Leonard and Lotka-Volterra is of growing interest in ecological modeling. Dynamical system exhibit chaotic behavior due to non-linear interaction and diffusion rules. The particular subject of chaotic properties of these models has been of growing interest and importance in the study of systems’ biodiversity. Our contribution relies on the use of statistical analysis to quantify the Lyapunov exponent for an increasing number of species in each May-Leonard and Lotka-Volterra models. Starting from slightly different initial conditions, Hamming distance curves are obtained and we use the \textit{q}-exponential function to find adequate fitting parameters. The convergence of Lyapunov exponents occurs as we increase the number of species.}

\section{Introduction}

In the last two decades, biodiversity has become a central topic of interest in nonlinear science in general. Due to its specific features and importance, biodiversity stands as a central and complex subject which plays an essential role to the viability of ecological systems. In particular, the diversity of species in ecosystems is known to be promoted by cyclic, non-hierarchical interactions among the competing populations \cite{kerr2002local, kirkup2004antibiotic, reichenbach2007mobility, Liao}. The mathematical field that attempts to formally describe such relational characteristics involving competing strategies, is known as game theory; see, for instance, \cite{von1959theory}. Interestingly, basic features of the non-transitive relations may be represented by the rules of the rock–paper–scissors game, in which rock breaks scissors, scissors cuts paper and paper covers rock, known as the RPS game. The subject is of current interest, and here we quote, for instance, the book \cite{nowak2006evolutionary}, which deals with several important aspects of evolutionary dynamics and two recent reviews, one dealing with spatial pattern formation, focusing on the spatiotemporal dynamics and on self-organization of socially favorable states \cite{perc2017statistical}, and the other concerned with different aspects of biodiversity, with focus on how it can be maintained under the rules of mathematical modeling \cite{Rev2}.

One basic feature, regarding the interaction among species in an ecosystem, is that the presence of the non-hierarchical cyclic rules of the RPS game enables competing species to mutually inhibit each others population. Due to the unstructured predation chain, the whole system evolves to a stable dynamics, showing some interesting spatial features{\color{black}. Among them, } the appearance of spiral patterns emerging from very simple non-hierarchical rules, as will be explicitly shown below. A particularly interesting result known as the survival of the weakest was obtained in \cite{frean2001rock}. It has unveiled that under specific asymmetry in the RPS rules, the weakening of a specific species may increase its presence in the system. This result has been confirmed in several other investigations, in particular, in the recent study using strains of \textit{Escherichia Coli} in Ref. \cite{Liao}. There, the authors have shown that intrinsic differences in the three major mechanisms of bacterial warfare may change the dynamics and conduct the system to an unbalanced community that is dominated by the weakest strain, in accordance with the result of Ref. \cite{frean2001rock}.

An important characteristic that appears under {\color{black} the} simple non-hierarchical rules of the RPS game, is the formation of spiral patterns. In this sense, it is of current interest to remember that spiral patterns are also present in several other system, for instance, in the human brain as spiral-like rotational wave patterns that spring abundantly during both resting and cognitive task states \cite{xu2023interacting}. Moreover, spiral structures can also appear as topological defects on the oocyte membranes of the starfish {\it{Patiria miniata}}. This has been studied recently in \cite{tan2020topological} and an interesting perspective would concern the possibility to shed more light on direct connections between spiral patterns and the dynamics of topological defect in living systems.

The basic features of biodiversity described with the simple RPS rules have also been studied in the last decades under several other perspectives, for example, changing the rules, adding new rules and enlarging the number of species, to quote some specific possibilities \cite{perc2017statistical, Rev2}. In a more recent work \cite{bazeiascirep}, some of us have developed a tool to qualify the chaotic behavior for a three-species model of biodiversity, with competition being driven by the cyclic rules of the RPS game. The investigation was based on the Hamming Distance Density (HDD), and the results have soon been generalized in \cite{bazeia2017hamming} for more species, including up to ten distinct species. Historically, the Hamming Distance (HD) was introduced in \cite{hamming1950error} as a way do distinguish quantities such as vectors, matrices, etc. For instance, the HD between the two following vectors $(0, 0, 1)$ and $(0, 1, 1)$ is one, since they differ from one another by a single digit. The HDD was studied in \cite{bazeiascirep} in the context of biodiversity, under the perspective that the square lattice used to simulate species in evolution can be seen as a square matrix. In this sense, the method of acquiring the desired curves reduces to the counting of distinct elements in two square lattices that evolve under the same rules, leading to a procedure rooted in solid ground. The HDD exhibits a universal profile uncovered in \cite{bazeiascirep} and confirmed in \cite{bazeia2017hamming}, that resembles a sigmoid function. It describes an initial evolution of an exponential-like growth and is then followed by a saturation plateau, which depends on the number of species.

In another direction, in the recent work \cite{avelino2022lotka} the authors studied the dynamical evolution of species and the pattern formation in two distinct cases, one considering the Lotka-Volterra (LV) \cite{lotka1920analytical, bacaer2011lotka} and the other the May-Leonard (ML) \cite{may1975nonlinear} algorithms introduced in the predation rule driven by the RPS game. More recently, in \cite{csf23} the authors considered a model with seven distinct species, investigating the possibility of transitions from seven to five species, and from five to three, and then from three to one species. The transitions are driven by the increase in the rule of mobility, leading to the cascading effect. This identification is in accordance with Ref. \cite{reichenbach2007mobility}, which showed that the increase of mobility may jeopardize biodiversity. In direct connection, in a recent work \cite{REV} the authors provide a brief review of the role played by phase transitions in explaining processes of biological interest. In another very recent work \cite{BO}, some of us have investigated the possibility to study the chaotic behavior in models of the public goods game type using the method of the \textit{q}-exponential. The results have encouraged us to go further on in the procedure, to study quantitative chaotic behavior in models described by the rules of the RPS game, considering the two possibilities of LV and ML evolutions. In order to enhance the power of the investigation, we have also considered several distinct systems, with the number $S$ of species increasing from $3$ to $12$, with the predation rule adapted to include interaction with second (and third, fourth, etc) neighbors as explained in the next Sec. \ref{sec2}. Since the Hamming distance follows an almost exponential growth at the beginning of the dynamical evolution, we use this profile to be fitted by the \textit{q}-exponential function, from where we extract the Lyapunov exponent corresponding to the dynamical evolution in both LV and ML cases. 

In order to implement the investigation, we opted to organize the present study as follows: In Sec. \ref{sec2} we describe the models to be considered in this work. There, we deal with the number of species and the rules of mobility, reproduction and predation, and the way they are implemented in the numerical simulations, for both approaches. The results are then presented in Sec. \ref{sec3}, where we also detail the main aspects of the algorithms used in the simulations. In particular, we discuss the chaotic aspects of each scenario via the fitting parameters. As another result of interest, we also study the spatial autocorrelation and calculate the correlation length in terms of the number of species in each family of systems. We end the work in Sec. \ref{end}, adding a summary of the main results and suggesting future developments in the subject.

\section{Methodology}
\label{sec2}

Our approach towards investigating different aspects of chaotic properties involves exploring and comparing two models of population dynamics, as we increase the number of competing species in the ecosystem \cite{szabo2008phase, Ave1, Ave2, vukov2013diverging}. In particular, we investigate the evolution of LV and ML systems and analyze quantitative and qualitative differences between such approaches. Both models are based on the same principles. We use a discrete lattice space with periodic boundary conditions, a discrete time evolution and a set of rules to be performed by the players in the simulations. Also, every action that is carried out on simulations are of random nature. Such simulations are known as cellular automata (CA), see \cite{neumann1966theory, ilachinski2001cellular}.
The lattice is composed of $N\times N$ identical square sites. The individuals occupying each site has a chance to interact with its four nearest neighbors (up, down, left and right). In the spatial lattice one considers periodic boundary conditions. In order to avoid finiteness problems with the lattice, we shall only deal with very large values of $N$.

\begin{figure}[ht]
	\centering
	\includegraphics[width=8.4cm]{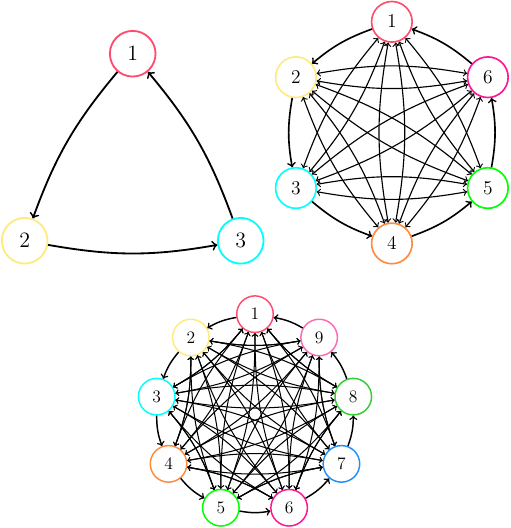}
	\caption{Illustration of the non-hierarchical predation chain for 3, 6 and 9 competing species. Interaction between the closest neighboring individuals is unidirectional, but bidirectional for all the other cases.}
	\label{fig:1}
\end{figure}

\begin{figure}[ht]
	\centering
	\includegraphics[width=8.4cm]{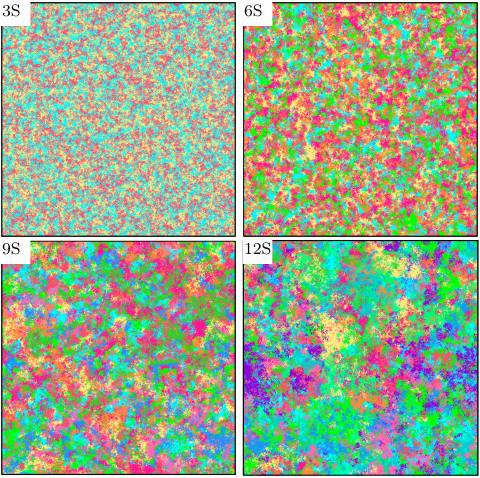}
	\caption{Typical final states of a single run simulation for $3$, $6$, $9$ and $12$ species in the LV model. The lattice size is $10^3\times 10^3$.}
	\label{fig:2}
\end{figure}

\begin{figure}[htbp]
	\centering
	\includegraphics[width=8.4cm]{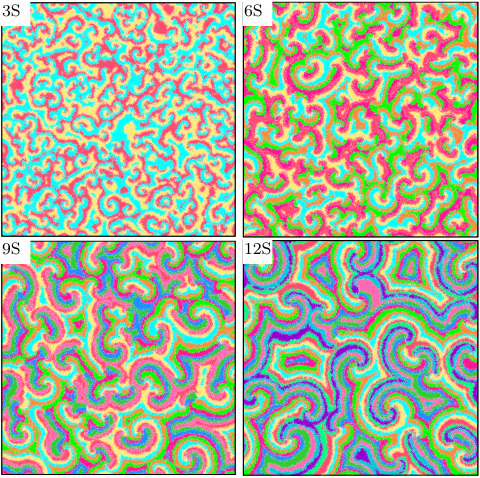}
	\caption{Typical final states of a single run simulation for $3$, $6$, $9$ and $12$ species in the ML model. The lattice size is $10^3\times 10^3$.}
	\label{fig:3}
\end{figure}

{As one knows, CA simulations are widely used for studying dynamical systems \cite{ermentrout1993cellular}. In the area of biodiversity, relevant results were already obtained with numerical simulations of this kind, as reported in \cite{nowak1992evolutionary, ermentrout1993cellular, kerr2002local}, as well as more recent investigations such as in Refs. \cite{bazeiascirep, bazeia2017hamming, alfaro2024hamming}, concerned with the phenomenology of population dynamics and the development of new computational techniques. In the present work, } our investigation enlarges the number of species to study systems with $3$ species up to $12$, for the two distinct cases of ML and LV dynamics. The species compete among each other in a non-hierarchical way, as illustrated in Fig. \ref{fig:1}. As we can see, interactions among closest neighboring individuals are unidirectional, and the others, among next, next-to-next etc neighbors are all bidirectional. We also display representative images of the state of the systems in Fig. \ref{fig:2} and Fig. \ref{fig:3} and there one sees the presence of spiral patterns only in the case of ML dynamics, as expected.

The algorithms used in this work encompass two distinct evolution. Basically, they differ in the implementation of the interaction rule. In the LV algorithm, one participating individual can either engage {\it mobility} $(m)$ or {\it predation} $(p)$, with probabilities $m$ and $p$ such that $m+p=1$. In this case, when the predator annihilates its prey, it simultaneously reproduces, leaving a copy of it at the prey site. On the other side, in ML context, the predation rule is split into two actions. In this case, an individual can execute {\it mobility} $(m)$, {\it reproduction} ($r$) or {\it predation} $(p)$, such that $m+r+p=1$. Here, predation works just to annihilates the prey, leaving an empty site. While in reproduction, an individual is able to make a copy of itself on an empty site.

In general, the model works as follows: first one initial state is created by randomly distributing individuals of each species and the empty spaces, if it is the case, throughout the sites of the discrete square lattice. This, therefore, creates an approximately homogeneous initial state of the system. The dynamical evolution begins via another random process. One site is chosen by chance, this is the active one. It is going to interact, or not, with one of the four nearest adjacent sites around it (up, down, left and right). As one may expect, the interacting neighbor site is selected by chance too, and it is labeled as the passive site. In sequence, the computer draws one of the available actions. For LV, the actions are $(m, p)$ and for ML, $(m, r, p)$. If the action selected is mobility $(m)$ the individuals in the active and passive sites are simply exchanged. This rule is identical on both models LV and ML. It is an action that promotes diffusion in the ecosystem. 

As mentioned before, in the ML scenario one has two additional rules. If the rule selected is reproduction, and if the neighboring site is empty, the individual selected is capable of reproducing itself with probability $r$. In this way it has the capacity of increasing its population. The third rule allows the different populations to interact with themselves. It is the predation rule $p$. If this rule is chosen and the passive site is inhabited by one of the preys according to the predation chain, see Fig. \ref{fig:1}, its site is left empty. On the other hand, in the LV model the reproduction and the predation actions are welded together. In this way, the predatory behavior occurs by replacing the prey site with an identical copy of itself. This enables a fixed number of individuals throughout the whole evolution. The same is not true for ML situation, where the total number of individual fluctuates through time.

After $N^2$ draws, we count a time unity, namely, a generation, or a Monte Carlo (MC) time unity.

\section{Results}
\label{sec3}

Let us now study the two families of models, with ML and LV rules, in the case of $3, 4, ..., 11$ and $12$ species. We first focus on the chaotic behavior of the systems, and then we deal with the correlation length in the same cases.

\subsection{Chaotic behavior}

\begin{figure}[ht]
	\centering
	\includegraphics[width=8.4cm]{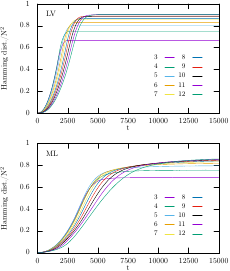}
	\caption{The Hamming distance density for $S=3, 4, \dots, 12$. Each curve is averaged over $10^3$ simulations realized on a lattice of size $10^3\times 10^3$. The top panel is for LV and the bottom panel for ML.} 
	\label{fig:4}
\end{figure}

\begin{figure}[ht]
	\centering
	\includegraphics[width=8.4cm]{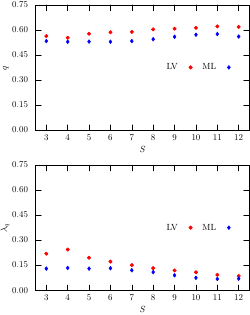}
	\caption{Data obtained from the \textit{q}-exponential fitting of the HDD for an increasing number of species in LV and ML models. The values of $q$ {\color{black} are} displayed on the top panel. The values of the coefficient $\lambda_q$ {\color{black} are} displayed on the bottom panel. Each point is averaged over $10^3$ simulations realized on a lattice of size $10^3\times 10^3$.}
	\label{fig:5}
\end{figure}

The main purpose here is to investigate the chaotic behavior of the ML and LV evolutions. We will implement this possibility using the Hamming Distance Density (HDD) approach. From previous works, one sees that the initial growth of the HDD follows a quasi-exponential profile. We compute the HDD following the lines of Refs. \cite{bazeiascirep, bazeia2017hamming}. The idea is inspired on the butterfly effect, that is, the sensitivity of the system to small changes on the initial conditions, usually present in chaotic systems. In order to identify this property we run two slightly different simulations in parallel. In detail, an initial state is generated and stored in the computer's memory. We then make an identical copy of it, but in this copy a random change of a single site is executed. With these two slightly different initial states, we made evolve the simulation applying the same rules on both lattices. It is important to stress out that the algorithm attempts to apply the same rule in both lattices regarding the neighborhood of each case. At the end of every generation, we compute the difference between the simulations via the Hamming Distance (HD) metric. We compare the occupation on each site of the lattices one by one. And if the occupations are different, one unity is summed up, otherwise nothing is registered and we follow the site scanning process. Finally, the resulting number is divided by the total area of the ecosystem to get the desired density. The procedure leads to the HDD. Each curve is obtained averaging over $10^3$ simulations realized on a lattice of size $10^3\times 10^3$. {\color{black} Throughout the paper, the standard error from the sample data is of order $10^{-4}$. Therefore, in Figs. \ref{fig:4} and \ref{fig:6} the uncertainties are within the line width, and in Figs. \ref{fig:5} and \ref{fig:7} they lie inside the size of the points.}

The simulations are implemented for every model, and the results are displayed in Fig. \ref{fig:4}. We notice that the HDD has the universal profile first identified in \cite{bazeiascirep} and soon after confirmed in \cite{bazeia2017hamming}. It must be emphasized here that the novelty is the calculation of the HDD for all the LV systems. Interestingly, it also keeps the universal profile of starting increasing almost exponentially and then saturating at some value that depends on the number of species in the system. We see from the HDD evolutions depicted in Fig. \ref{fig:4}, that for LV they relax to the saturation values faster than in the case of the ML systems. We think there is a natural interpretation for this: in the case of ML simulation one has three distinct actions; for LV, otherwise, one has two actions. This leads the LV simulations to evolve faster than in the ML case.

We see that in every case, the {HDD} curve has a similar profile. First, one notes a rapid growth, the difference between the two fields spreads in an almost exponential form for small values of $t$. As $t$ increases to larger and larger values, it acquires an asymptotic behavior, where the difference between the two simulations reaches a maximum value.

We highlight that there exists a species dependency in the asymptotic values of LV's HDD curves. When the differences between the systems, reaches a stationary state, one can interpret the HDD as the probability of checking different individuals occupying the very same site, on both lattices. Due to the homogeneity of the LV spatial patterns these numbers follow a well defined rule. Since $S$ is the number of species present in the simulation of interest, the difference between the two lattices becomes $(S-1)/S$. This means that, if we simultaneously pick the same random site on both lattices the chances they are equal is $1/S$. For example, if $S=3$, one third will be the probability of identical measures of this sort. While two thirds of the measures will result in mismatching. This same relation is not verified for ML curves since here, the presence of spatial structures and empty spaces does not lead to a clear straightforward rule.

{\color{black} The main innovation of this investigation lies in the study to fit the acquired curves using a \textit{q}-exponential function, as the authors of Ref. \cite{BO} have investigated. The motivation behind it relies on the fact that, chaotic systems exhibit an exponential separation of initially infinitesimally close trajectories in phase space. And this characteristic is quantified via the (positive) Lyapunov exponent $\lambda$ \cite{pikovsky2016lyapunov}, defined as

\begin{equation}
	\Delta H(t) \sim \Delta H(0) \exp(\lambda t) 
\end{equation}
for $\Delta H(0) \rightarrow 0$ and $t \rightarrow \infty$. 
Since in some class of phenomena, slightly different initial conditions deviate from each other in an almost exponential manner, such systems are called weakly and strongly chaotic systems. Thus, we introduce $\lambda_q$, with the sub-index $q$ used to identify these possibilities. In this case, $\lambda_q< 1$ and $\lambda_q > 1$ identify the weakly and strongly chaotic cases, respectively \cite{tsallis1997power, tirnakli2016standard}. The presence of chaos in the system is then described via the \textit{q}-exponential equation bellow}

\begin{equation}
	\Delta H_q(t)\sim \Delta H_q(0)\;(1+(1-q)\lambda_q t)^{[1/(1-q)]}
	\label{eq:1}.
\end{equation}

In the limit $q\to1$, it engenders the exponential profile $e^{\lambda t}$, where $\lambda$ ($\lambda_q\to\lambda_1=\lambda$) represents the standard Lyapunov exponent. In Fig. \ref{fig:5} we depict the results obtained for both $q$ and $\lambda_q$, when one uses Eq. \eqref{eq:1} to fit the HDD. This is done using a gnuplot based fitting function, documented in \cite{gnuplot-doc}. 

As one can see from the results displayed in Fig. \ref{fig:5}, the values of $q$ remain practically constant for both cases, independently of the number of species. It is around $0.6$ for LV and a little less for ML evolutions. The parameter $\lambda_q$, however, has another behavior; they are well distinct from one another, and one notices a slight tendency to decrease as $S$ increases to larger values. In this figure, in both LV and ML evolutions, each point is computed at the time when the difference between the two lattices is $1\%$. For the initial state prepared at random disposition of the possible occupations, we see that the values of $\lambda_q$ stay slightly higher in LV models in comparison with ML. This indicates that two slightly different trajectories separate faster under the context of the LV rules.

\subsection{Correlation length}

Let us now pay attention to the calculation of the correlation length for the two families of models. We follow the procedure utilized in \cite{avelino2022lotka}. This quantity allow us to infer the average size of the patches, or domains, made of clusters of the same species that forms in the system. As one sees, in Fig. \ref{fig:2} in LV simulations every species is able to occupy a region with unclear borders or regular shapes. The groups of same individuals, neighbors each other according to its predatory preferences. On the other hand, in the ML case, Fig. \ref{fig:3}, the appearance of spiral patterns is visible. These structures indicates a stronger spatial correlations between individuals and its surrounding.

To quantify this spatial property, we compute the discrete spatial autocorrelation function, $C(r)$. We define the scalar field $\phi_{i, j}$ that represents individuals in the position $(i, j)$ at the lattice. Then the autocorrelation function is defined by 

\begin{equation}
	C(r) = \sum_{m, n \in S(r)} \dfrac{f_{m, n}}{\mu f_{0, 0}} \ , 
	\label{eq1}
\end{equation} 
where
\begin{equation}
	S(r)=\{(i, j) \in \mathbb{N} : i+j= \frac{r}{\Delta r} \cap i \leq N \cap j< N \} \, 
\end{equation}
is the region consisting of the sites available in the lattice. Here $\cap$ stands for the \textit{intersection} operation among the groups involved, and it limits the sum $i+j= r/\Delta r$ up to the edges of the system. Moreover, 

\begin{equation}
	\mu=\mathrm{min}[2N-(m+n+1), m+n+1], 
\end{equation}
is a function that computes the minimum between the comma separated values, and

\begin{equation}
f_{m, n}= \sum_{i=1}^{N}\sum_{j=1}^{N}\varphi_{i, j}\varphi_{i+m, n+l} .
\end{equation}

Here we take $\Delta r = 1$ as the spacing between the lattice sites and each species individual being represented by $\varphi_{i, j}=\phi_{i, j}-\Bar{\phi}$, with $\Bar{\phi}$ as the mean value of $\phi$. It is important to note that, $C(r)$ is defined only on the lattice, \textit{i.e.} for $\{r\in \mathcal{A}\, | \, \mathcal{A}=\mathbb{N}\cap r \leq 2N\}$.

\begin{figure}[ht]
	\centering
	\includegraphics[width=8.4cm]{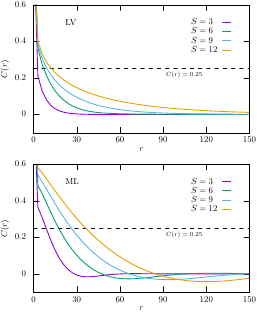}
	\caption{Spatial autocorrelation function for the LV and ML models. The horizontal black line represents the characteristic length $\ell$. We compute each curve on a representative state of a $2\cdot 10^4\times 2\cdot 10^4$ lattice at $t=5\cdot 10^3$.}
	\label{fig:6}
\end{figure}

The profile of $C(r)$ is displayed in Fig. \ref{fig:6} in the cases of LV and ML interactions, for 3, 6, 9, and 12 species. The curves are obtained at the final state of a single numerical simulation in a lattice of size $2\cdot 10^4\times 2\cdot 10^4$, at time $t=5\cdot 10^3$. 

\begin{figure}[ht]
	\centering
	\includegraphics[width=8.4cm]{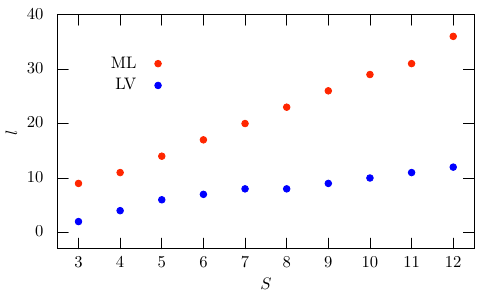}
	\caption{Characteristic length $\ell$ as a function of the number of species $S$. Each point is obtained from the respective autocorrelation function when $C(r)=0.25$. The measurements occur on the final state of a single run simulation on a lattice size $2\cdot 10^4\times 2\cdot 10^4$ at time $t=5\cdot 10^3$.}
	\label{fig:7}
\end{figure}

We go on and calculate the characteristic length, $\ell$, which is obtained at $C(r)=0.25$, from where we register the intersections. The results are depicted in Fig. \ref{fig:7} for $S=3$, $4$, $\cdots, 12$, in both ML and LV cases. Each point is obtained from the respective autocorrelation function when $C(r)=0.25$. The measurements occur on the final state of a single run simulation on a lattice size $2\cdot 10^4\times 2\cdot 10^4$, at time $t=5\cdot 10^3$. We notice that the characteristic length grows following an almost linear tendency while $S$ also grows. This holds stronger for the ML case. As displayed in Fig. \ref{fig:7}, ML simulations have larger $\ell$ values when compared with LV, which is in accordance with the qualitative results shown in Figs. \ref{fig:2} and \ref{fig:3}, and also, with the investigation of chaotic behavior studied in Sec. III.A.

\section{Ending comments}
\label{end}

In this work we have investigated chaotic properties of generalized RPS dynamics under non-hierarchical interactions of both the LV and ML types. We approached the subject considering the Hamming distance density, which we calculated for models with several species, $S$, considering the cases $S=3, 4, $ ... $, 12$. We did the simulations using the two distinct dynamics, with the LV and ML interactions. The results for the LV models are all original, calculated here for the first time. And they corroborate the suggestion introduced in \cite{bazeiascirep, bazeia2017hamming}, that the Hamming distance follows universal behavior, always increasing almost exponential initially, and then saturating at some specific value which depends on the number of species in the system.

Inspired by the behavior of the Hamming distance, we then fitted its profile with the \textit{q}-exponential function, which is an almost exponential that recuperates the exponential behavior in the limit $q\to1$. We used the fitting to infer the Lyapunov exponent for the two LV and ML dynamics. Interestingly, we found that the exponents are only slightly different, but they are practically independent of the number of species. We also noted that in the case of LV interactions, it is always a little larger then in the case of ML. This is another novelty, which is in accordance with the fact that the LV simulations evolve faster than the ML, because the LV rules requires a two-step implementation, with mobility and predation, while the ML rules needs a three-step implementation, using mobility, predation and reproduction. 

We have also investigated the spatial autocorrelation length and the characteristic length for the two families of systems. We found interesting results, with the characteristic length, in particular, being higher for ML interactions, but both showing an almost linear increasing when one increases the number of species. These results are in accordance with the qualitative results depicted in Fig. \ref{fig:2} for LV systems and in Fig. \ref{fig:3} for ML systems.

The present results reinforce the point of view that the dynamical evolution of species governed by simple non-hierarchical rules of the RPS game engenders chaotic behavior. They also reveal interesting differences in the chaotic properties of the dynamical evolution under the {\it Lotka-Volterra} and the {\it May-Leonard} interactions, and we expect to continue these investigations in other distinct ways, exploring the effects of a local carrying capacity parameter that may be softly varied in the environment, as studied in \cite{bazeia2021environment}, and modifying the predation chain and strategy, as done in \cite{bazeia2021effects}. We also intend to study how the chaotic behavior may depend on the number of species, when the rules are changed to induce the cascading transitions identified in \cite{csf23}. Another issue concerns the influence of the lattice characteristics, in particular, under modification of the topology of the environment where the species evolve.

\begin{center}{*****}\end{center}

This work is supported in part by CAPES (Grant 88887.606927/2021-00), CNPq (Grants nos. 404913/2018-0, 303469/2019-6 and 309835/2022-4), Paraíba State Research Foundation (FAPESQ-PB, Grant no. 0015/2019), and by Fundação Araucária and INCT-FCx (CNPq/FAPESP).


\begin{thebibliography}{36}%
\makeatletter
\providecommand \@ifxundefined [1]{%
 \@ifx{#1\undefined}
}%
\providecommand \@ifnum [1]{%
 \ifnum #1\expandafter \@firstoftwo
 \else \expandafter \@secondoftwo
 \fi
}%
\providecommand \@ifx [1]{%
 \ifx #1\expandafter \@firstoftwo
 \else \expandafter \@secondoftwo
 \fi
}%
\providecommand \natexlab [1]{#1}%
\providecommand \enquote  [1]{``#1''}%
\providecommand \bibnamefont  [1]{#1}%
\providecommand \bibfnamefont [1]{#1}%
\providecommand \citenamefont [1]{#1}%
\providecommand \href@noop [0]{\@secondoftwo}%
\providecommand \href [0]{\begingroup \@sanitize@url \@href}%
\providecommand \@href[1]{\@@startlink{#1}\@@href}%
\providecommand \@@href[1]{\endgroup#1\@@endlink}%
\providecommand \@sanitize@url [0]{\catcode `\\12\catcode `\$12\catcode
  `\&12\catcode `\#12\catcode `\^12\catcode `\_12\catcode `\%12\relax}%
\providecommand \@@startlink[1]{}%
\providecommand \@@endlink[0]{}%
\providecommand \url  [0]{\begingroup\@sanitize@url \@url }%
\providecommand \@url [1]{\endgroup\@href {#1}{\urlprefix }}%
\providecommand \urlprefix  [0]{URL }%
\providecommand \Eprint [0]{\href }%
\providecommand \doibase [0]{https://doi.org/}%
\providecommand \selectlanguage [0]{\@gobble}%
\providecommand \bibinfo  [0]{\@secondoftwo}%
\providecommand \bibfield  [0]{\@secondoftwo}%
\providecommand \translation [1]{[#1]}%
\providecommand \BibitemOpen [0]{}%
\providecommand \bibitemStop [0]{}%
\providecommand \bibitemNoStop [0]{.\EOS\space}%
\providecommand \EOS [0]{\spacefactor3000\relax}%
\providecommand \BibitemShut  [1]{\csname bibitem#1\endcsname}%
\let\auto@bib@innerbib\@empty
%</preamble>
{\color{black}
\bibitem [{\citenamefont {Kerr}\ \emph {et~al.}(2002)\citenamefont {Kerr},
  \citenamefont {Riley}, \citenamefont {Feldman},\ and\ \citenamefont
  {Bohannan}}]{kerr2002local}%
  \BibitemOpen
  \bibfield  {author} {\bibinfo {author} {\bibfnamefont {B.}~\bibnamefont
  {Kerr}}, \bibinfo {author} {\bibfnamefont {M.~A.}\ \bibnamefont {Riley}},
  \bibinfo {author} {\bibfnamefont {M.~W.}\ \bibnamefont {Feldman}},\ and\
  \bibinfo {author} {\bibfnamefont {B.~J.~M.}\ \bibnamefont {Bohannan}},\
  }\href {https://doi.org/10.1038/nature00823} {\bibfield  {journal} {\bibinfo
  {journal} {Nature}\ }\textbf {\bibinfo {volume} {418}},\ \bibinfo {pages}
  {171–174} (\bibinfo {year} {2002})}\BibitemShut {NoStop}%
\bibitem [{\citenamefont {Kirkup}\ and\ \citenamefont
  {Riley}(2004)}]{kirkup2004antibiotic}%
  \BibitemOpen
  \bibfield  {author} {\bibinfo {author} {\bibfnamefont {B.~C.}\ \bibnamefont
  {Kirkup}}\ and\ \bibinfo {author} {\bibfnamefont {M.~A.}\ \bibnamefont
  {Riley}},\ }\href {https://doi.org/10.1038/nature02429} {\bibfield  {journal}
  {\bibinfo  {journal} {Nature}\ }\textbf {\bibinfo {volume} {428}},\ \bibinfo
  {pages} {412–414} (\bibinfo {year} {2004})}\BibitemShut {NoStop}%
\bibitem [{\citenamefont {Reichenbach}\ \emph {et~al.}(2007)\citenamefont
  {Reichenbach}, \citenamefont {Mobilia},\ and\ \citenamefont
  {Frey}}]{reichenbach2007mobility}%
  \BibitemOpen
  \bibfield  {author} {\bibinfo {author} {\bibfnamefont {T.}~\bibnamefont
  {Reichenbach}}, \bibinfo {author} {\bibfnamefont {M.}~\bibnamefont
  {Mobilia}},\ and\ \bibinfo {author} {\bibfnamefont {E.}~\bibnamefont
  {Frey}},\ }\href {https://doi.org/10.1038/nature06095} {\bibfield  {journal}
  {\bibinfo  {journal} {Nature}\ }\textbf {\bibinfo {volume} {448}},\ \bibinfo
  {pages} {1046–1049} (\bibinfo {year} {2007})}\BibitemShut {NoStop}%
\bibitem [{\citenamefont {Liao}\ \emph {et~al.}(2020)\citenamefont {Liao},
  \citenamefont {Miano}, \citenamefont {Nguyen}, \citenamefont {Chao},\ and\
  \citenamefont {Hasty}}]{Liao}%
  \BibitemOpen
  \bibfield  {author} {\bibinfo {author} {\bibfnamefont {M.~J.}\ \bibnamefont
  {Liao}}, \bibinfo {author} {\bibfnamefont {A.}~\bibnamefont {Miano}},
  \bibinfo {author} {\bibfnamefont {C.~B.}\ \bibnamefont {Nguyen}}, \bibinfo
  {author} {\bibfnamefont {L.}~\bibnamefont {Chao}},\ and\ \bibinfo {author}
  {\bibfnamefont {J.}~\bibnamefont {Hasty}},\ }\href
  {https://doi.org/10.1038/s41467-020-19963-8} {\bibfield  {journal} {\bibinfo
  {journal} {Nature Communications}\ }\textbf {\bibinfo {volume} {11}},\
  \bibinfo {pages} {6055} (\bibinfo {year} {2020})}\BibitemShut {NoStop}%
\bibitem [{\citenamefont {von Neumann}(1959)}]{von1959theory}%
  \BibitemOpen
  \bibfield  {author} {\bibinfo {author} {\bibfnamefont {J.}~\bibnamefont {von
  Neumann}},\ }\bibinfo {title} {1. on the theory of games of strategy},\ in\
  \href {https://doi.org/10.1515/9781400882168-003} {\emph {\bibinfo
  {booktitle} {Contributions to the Theory of Games (AM-40), Volume IV}}},\
  \bibinfo {editor} {edited by\ \bibinfo {editor} {\bibfnamefont {A.~W.}\
  \bibnamefont {Tucker}}\ and\ \bibinfo {editor} {\bibfnamefont {R.~D.}\
  \bibnamefont {Luce}}}\ (\bibinfo  {publisher} {Princeton University Press},\
  \bibinfo {address} {Princeton},\ \bibinfo {year} {1959})\ pp.\ \bibinfo
  {pages} {13--42}\BibitemShut {NoStop}%
\bibitem [{\citenamefont {Nowak}(2006)}]{nowak2006evolutionary}%
  \BibitemOpen
  \bibfield  {author} {\bibinfo {author} {\bibfnamefont {M.}~\bibnamefont
  {Nowak}},\ }\href@noop {} {\emph {\bibinfo {title} {Evolutionary Dynamics:
  Exploring the Equations of Life}}}\ (\bibinfo  {publisher} {Harvard
  University Press},\ \bibinfo {address} {no},\ \bibinfo {year}
  {2006})\BibitemShut {NoStop}%
\bibitem [{\citenamefont {Perc}\ \emph {et~al.}(2017)\citenamefont {Perc},
  \citenamefont {Jordan}, \citenamefont {Rand}, \citenamefont {Wang},
  \citenamefont {Boccaletti},\ and\ \citenamefont
  {Szolnoki}}]{perc2017statistical}%
  \BibitemOpen
  \bibfield  {author} {\bibinfo {author} {\bibfnamefont {M.}~\bibnamefont
  {Perc}}, \bibinfo {author} {\bibfnamefont {J.~J.}\ \bibnamefont {Jordan}},
  \bibinfo {author} {\bibfnamefont {D.~G.}\ \bibnamefont {Rand}}, \bibinfo
  {author} {\bibfnamefont {Z.}~\bibnamefont {Wang}}, \bibinfo {author}
  {\bibfnamefont {S.}~\bibnamefont {Boccaletti}},\ and\ \bibinfo {author}
  {\bibfnamefont {A.}~\bibnamefont {Szolnoki}},\ }\href
  {https://doi.org/10.1016/j.physrep.2017.05.004} {\bibfield  {journal}
  {\bibinfo  {journal} {Physics Reports}\ }\textbf {\bibinfo {volume} {687}},\
  \bibinfo {pages} {1–51} (\bibinfo {year} {2017})}\BibitemShut {NoStop}%
\bibitem [{\citenamefont {Szolnoki}\ \emph {et~al.}(2020)\citenamefont
  {Szolnoki}, \citenamefont {de~Oliveira},\ and\ \citenamefont
  {Bazeia}}]{Rev2}%
  \BibitemOpen
  \bibfield  {author} {\bibinfo {author} {\bibfnamefont {A.}~\bibnamefont
  {Szolnoki}}, \bibinfo {author} {\bibfnamefont {B.~F.}\ \bibnamefont
  {de~Oliveira}},\ and\ \bibinfo {author} {\bibfnamefont {D.}~\bibnamefont
  {Bazeia}},\ }\href {https://doi.org/10.1209/0295-5075/131/68001} {\bibfield
  {journal} {\bibinfo  {journal} {Europhysics Letters}\ }\textbf {\bibinfo
  {volume} {131}},\ \bibinfo {pages} {68001} (\bibinfo {year}
  {2020})}\BibitemShut {NoStop}%
\bibitem [{\citenamefont {Frean}\ and\ \citenamefont
  {Abraham}(2001)}]{frean2001rock}%
  \BibitemOpen
  \bibfield  {author} {\bibinfo {author} {\bibfnamefont {M.}~\bibnamefont
  {Frean}}\ and\ \bibinfo {author} {\bibfnamefont {E.~R.}\ \bibnamefont
  {Abraham}},\ }\href {https://doi.org/10.1098/rspb.2001.1670} {\bibfield
  {journal} {\bibinfo  {journal} {Proceedings of the Royal Society of London.
  Series B: Biological Sciences}\ }\textbf {\bibinfo {volume} {268}},\ \bibinfo
  {pages} {1323–1327} (\bibinfo {year} {2001})}\BibitemShut {NoStop}%
\bibitem [{\citenamefont {Xu}\ \emph {et~al.}(2023)\citenamefont {Xu},
  \citenamefont {Long}, \citenamefont {Feng},\ and\ \citenamefont
  {Gong}}]{xu2023interacting}%
  \BibitemOpen
  \bibfield  {author} {\bibinfo {author} {\bibfnamefont {Y.}~\bibnamefont
  {Xu}}, \bibinfo {author} {\bibfnamefont {X.}~\bibnamefont {Long}}, \bibinfo
  {author} {\bibfnamefont {J.}~\bibnamefont {Feng}},\ and\ \bibinfo {author}
  {\bibfnamefont {P.}~\bibnamefont {Gong}},\ }\href
  {https://doi.org/10.1038/s41562-023-01626-5} {\bibfield  {journal} {\bibinfo
  {journal} {Nature Human Behaviour}\ }\textbf {\bibinfo {volume} {7}},\
  \bibinfo {pages} {1196–1215} (\bibinfo {year} {2023})}\BibitemShut
  {NoStop}%
\bibitem [{\citenamefont {Tan}\ \emph {et~al.}(2020)\citenamefont {Tan},
  \citenamefont {Liu}, \citenamefont {Miller}, \citenamefont {Tekant},
  \citenamefont {Dunkel},\ and\ \citenamefont {Fakhri}}]{tan2020topological}%
  \BibitemOpen
  \bibfield  {author} {\bibinfo {author} {\bibfnamefont {T.~H.}\ \bibnamefont
  {Tan}}, \bibinfo {author} {\bibfnamefont {J.}~\bibnamefont {Liu}}, \bibinfo
  {author} {\bibfnamefont {P.~W.}\ \bibnamefont {Miller}}, \bibinfo {author}
  {\bibfnamefont {M.}~\bibnamefont {Tekant}}, \bibinfo {author} {\bibfnamefont
  {J.}~\bibnamefont {Dunkel}},\ and\ \bibinfo {author} {\bibfnamefont
  {N.}~\bibnamefont {Fakhri}},\ }\href
  {https://doi.org/10.1038/s41567-020-0841-9} {\bibfield  {journal} {\bibinfo
  {journal} {Nature Physics}\ }\textbf {\bibinfo {volume} {16}},\ \bibinfo
  {pages} {657–662} (\bibinfo {year} {2020})}\BibitemShut {NoStop}%
\bibitem [{\citenamefont {Bazeia}\ \emph
  {et~al.}(2017{\natexlab{a}})\citenamefont {Bazeia}, \citenamefont {Pereira},
  \citenamefont {Brito}, \citenamefont {Oliveira},\ and\ \citenamefont
  {Ramos}}]{bazeiascirep}%
  \BibitemOpen
  \bibfield  {author} {\bibinfo {author} {\bibfnamefont {D.}~\bibnamefont
  {Bazeia}}, \bibinfo {author} {\bibfnamefont {M.~B. P.~N.}\ \bibnamefont
  {Pereira}}, \bibinfo {author} {\bibfnamefont {A.~V.}\ \bibnamefont {Brito}},
  \bibinfo {author} {\bibfnamefont {B.~d.}\ \bibnamefont {Oliveira}},\ and\
  \bibinfo {author} {\bibfnamefont {J.~G. G.~S.}\ \bibnamefont {Ramos}},\
  }\href {https://doi.org/10.1038/srep44900} {\bibfield  {journal} {\bibinfo
  {journal} {Scientific Reports}\ }\textbf {\bibinfo {volume} {7}},\ \bibinfo
  {pages} {4490} (\bibinfo {year} {2017}{\natexlab{a}})}\BibitemShut {NoStop}%
\bibitem [{\citenamefont {Bazeia}\ \emph
  {et~al.}(2017{\natexlab{b}})\citenamefont {Bazeia}, \citenamefont {Menezes},
  \citenamefont {de~Oliveira},\ and\ \citenamefont
  {Ramos}}]{bazeia2017hamming}%
  \BibitemOpen
  \bibfield  {author} {\bibinfo {author} {\bibfnamefont {D.}~\bibnamefont
  {Bazeia}}, \bibinfo {author} {\bibfnamefont {J.}~\bibnamefont {Menezes}},
  \bibinfo {author} {\bibfnamefont {B.~F.}\ \bibnamefont {de~Oliveira}},\ and\
  \bibinfo {author} {\bibfnamefont {J.~G. G.~S.}\ \bibnamefont {Ramos}},\
  }\href {https://doi.org/10.1209/0295-5075/119/58003} {\bibfield  {journal}
  {\bibinfo  {journal} {Europhysics Letters}\ }\textbf {\bibinfo {volume}
  {119}},\ \bibinfo {pages} {58003} (\bibinfo {year}
  {2017}{\natexlab{b}})}\BibitemShut {NoStop}%
\bibitem [{\citenamefont {Hamming}(1950)}]{hamming1950error}%
  \BibitemOpen
  \bibfield  {author} {\bibinfo {author} {\bibfnamefont {R.~W.}\ \bibnamefont
  {Hamming}},\ }\href {https://doi.org/10.1002/j.1538-7305.1950.tb00463.x}
  {\bibfield  {journal} {\bibinfo  {journal} {Bell System Technical Journal}\
  }\textbf {\bibinfo {volume} {29}},\ \bibinfo {pages} {147–160} (\bibinfo
  {year} {1950})}\BibitemShut {NoStop}%
\bibitem [{\citenamefont {Avelino}\ \emph {et~al.}(2022)\citenamefont
  {Avelino}, \citenamefont {de~Oliveira},\ and\ \citenamefont
  {Trintin}}]{avelino2022lotka}%
  \BibitemOpen
  \bibfield  {author} {\bibinfo {author} {\bibfnamefont {P.~P.}\ \bibnamefont
  {Avelino}}, \bibinfo {author} {\bibfnamefont {B.~F.}\ \bibnamefont
  {de~Oliveira}},\ and\ \bibinfo {author} {\bibfnamefont {R.~S.}\ \bibnamefont
  {Trintin}},\ }\href {https://doi.org/10.1103/physreve.105.024309} {\bibfield
  {journal} {\bibinfo  {journal} {Physical Review E}\ }\textbf {\bibinfo
  {volume} {105}},\ \bibinfo {pages} {024309} (\bibinfo {year}
  {2022})}\BibitemShut {NoStop}%
\bibitem [{\citenamefont {Lotka}(1920)}]{lotka1920analytical}%
  \BibitemOpen
  \bibfield  {author} {\bibinfo {author} {\bibfnamefont {A.~J.}\ \bibnamefont
  {Lotka}},\ }\href {https://doi.org/10.1073/pnas.6.7.410} {\bibfield
  {journal} {\bibinfo  {journal} {Proceedings of the National Academy of
  Sciences}\ }\textbf {\bibinfo {volume} {6}},\ \bibinfo {pages} {410–415}
  (\bibinfo {year} {1920})}\BibitemShut {NoStop}%
\bibitem [{\citenamefont {Baca{\"e}r}(2011)}]{bacaer2011lotka}%
  \BibitemOpen
  \bibfield  {author} {\bibinfo {author} {\bibfnamefont {N.}~\bibnamefont
  {Baca{\"e}r}},\ }\bibinfo {title} {Lotka, volterra and the predator--prey
  system (1920--1926)},\ in\ \href
  {https://doi.org/10.1007/978-0-85729-115-8_13} {\emph {\bibinfo {booktitle}
  {A Short History of Mathematical Population Dynamics}}}\ (\bibinfo
  {publisher} {Springer London},\ \bibinfo {address} {London},\ \bibinfo {year}
  {2011})\ pp.\ \bibinfo {pages} {71--76}\BibitemShut {NoStop}%
\bibitem [{\citenamefont {May}\ and\ \citenamefont
  {Leonard}(1975)}]{may1975nonlinear}%
  \BibitemOpen
  \bibfield  {author} {\bibinfo {author} {\bibfnamefont {R.~M.}\ \bibnamefont
  {May}}\ and\ \bibinfo {author} {\bibfnamefont {W.~J.}\ \bibnamefont
  {Leonard}},\ }\href {https://doi.org/10.1137/0129022} {\bibfield  {journal}
  {\bibinfo  {journal} {SIAM Journal on Applied Mathematics}\ }\textbf
  {\bibinfo {volume} {29}},\ \bibinfo {pages} {243–253} (\bibinfo {year}
  {1975})}\BibitemShut {NoStop}%
\bibitem [{\citenamefont {Yang}\ and\ \citenamefont {Park}(2023)}]{csf23}%
  \BibitemOpen
  \bibfield  {author} {\bibinfo {author} {\bibfnamefont {R.~K.}\ \bibnamefont
  {Yang}}\ and\ \bibinfo {author} {\bibfnamefont {J.}~\bibnamefont {Park}},\
  }\href {https://doi.org/10.1016/j.chaos.2023.113949} {\bibfield  {journal}
  {\bibinfo  {journal} {Chaos, Solitons \& Fractals}\ }\textbf {\bibinfo
  {volume} {175}},\ \bibinfo {pages} {113949} (\bibinfo {year}
  {2023})}\BibitemShut {NoStop}%
\bibitem [{\citenamefont {Heffern}\ \emph {et~al.}(2021)\citenamefont
  {Heffern}, \citenamefont {Huelskamp}, \citenamefont {Bahar},\ and\
  \citenamefont {Inglis}}]{REV}%
  \BibitemOpen
  \bibfield  {author} {\bibinfo {author} {\bibfnamefont {E.~F.~W.}\
  \bibnamefont {Heffern}}, \bibinfo {author} {\bibfnamefont {H.}~\bibnamefont
  {Huelskamp}}, \bibinfo {author} {\bibfnamefont {S.}~\bibnamefont {Bahar}},\
  and\ \bibinfo {author} {\bibfnamefont {R.~F.}\ \bibnamefont {Inglis}},\
  }\href {https://doi.org/10.1098/rspb.2021.1111} {\bibfield  {journal}
  {\bibinfo  {journal} {Proceedings of the Royal Society B: Biological
  Sciences}\ }\textbf {\bibinfo {volume} {288}},\ \bibinfo {pages} {20211111}
  (\bibinfo {year} {2021})}\BibitemShut {NoStop}%
\bibitem [{\citenamefont {Bazeia}\ \emph {et~al.}(2023)\citenamefont {Bazeia},
  \citenamefont {Ferreira},\ and\ \citenamefont {de~Oliveira}}]{BO}%
  \BibitemOpen
  \bibfield  {author} {\bibinfo {author} {\bibfnamefont {D.}~\bibnamefont
  {Bazeia}}, \bibinfo {author} {\bibfnamefont {M.~J.~B.}\ \bibnamefont
  {Ferreira}},\ and\ \bibinfo {author} {\bibfnamefont {B.~F.}\ \bibnamefont
  {de~Oliveira}},\ }\href {https://doi.org/10.1209/0295-5075/ad01d9} {\bibfield
   {journal} {\bibinfo  {journal} {Europhysics Letters}\ }\textbf {\bibinfo
  {volume} {144}},\ \bibinfo {pages} {11007} (\bibinfo {year}
  {2023})}\BibitemShut {NoStop}%
\bibitem [{\citenamefont {Szabó}\ and\ \citenamefont
  {Szolnoki}(2008)}]{szabo2008phase}%
  \BibitemOpen
  \bibfield  {author} {\bibinfo {author} {\bibfnamefont {G.}~\bibnamefont
  {Szabó}}\ and\ \bibinfo {author} {\bibfnamefont {A.}~\bibnamefont
  {Szolnoki}},\ }\href {https://doi.org/10.1103/physreve.77.011906} {\bibfield
  {journal} {\bibinfo  {journal} {Physical Review E}\ }\textbf {\bibinfo
  {volume} {77}},\ \bibinfo {pages} {011906} (\bibinfo {year}
  {2008})}\BibitemShut {NoStop}%
\bibitem [{\citenamefont {Avelino}\ \emph
  {et~al.}(2012{\natexlab{a}})\citenamefont {Avelino}, \citenamefont {Bazeia},
  \citenamefont {Losano}, \citenamefont {Menezes},\ and\ \citenamefont
  {Oliveira}}]{Ave1}%
  \BibitemOpen
  \bibfield  {author} {\bibinfo {author} {\bibfnamefont {P.~P.}\ \bibnamefont
  {Avelino}}, \bibinfo {author} {\bibfnamefont {D.}~\bibnamefont {Bazeia}},
  \bibinfo {author} {\bibfnamefont {L.}~\bibnamefont {Losano}}, \bibinfo
  {author} {\bibfnamefont {J.}~\bibnamefont {Menezes}},\ and\ \bibinfo {author}
  {\bibfnamefont {B.~F.}\ \bibnamefont {Oliveira}},\ }\href
  {https://doi.org/10.1103/physreve.86.036112} {\bibfield  {journal} {\bibinfo
  {journal} {Physical Review E}\ }\textbf {\bibinfo {volume} {86}},\ \bibinfo
  {pages} {036112} (\bibinfo {year} {2012}{\natexlab{a}})}\BibitemShut
  {NoStop}%
\bibitem [{\citenamefont {Avelino}\ \emph
  {et~al.}(2012{\natexlab{b}})\citenamefont {Avelino}, \citenamefont {Bazeia},
  \citenamefont {Losano},\ and\ \citenamefont {Menezes}}]{Ave2}%
  \BibitemOpen
  \bibfield  {author} {\bibinfo {author} {\bibfnamefont {P.~P.}\ \bibnamefont
  {Avelino}}, \bibinfo {author} {\bibfnamefont {D.}~\bibnamefont {Bazeia}},
  \bibinfo {author} {\bibfnamefont {L.}~\bibnamefont {Losano}},\ and\ \bibinfo
  {author} {\bibfnamefont {J.}~\bibnamefont {Menezes}},\ }\href
  {https://doi.org/10.1103/physreve.86.031119} {\bibfield  {journal} {\bibinfo
  {journal} {Physical Review E}\ }\textbf {\bibinfo {volume} {86}},\ \bibinfo
  {pages} {031119} (\bibinfo {year} {2012}{\natexlab{b}})}\BibitemShut
  {NoStop}%
\bibitem [{\citenamefont {Vukov}\ \emph {et~al.}(2013)\citenamefont {Vukov},
  \citenamefont {Szolnoki},\ and\ \citenamefont {Szabó}}]{vukov2013diverging}%
  \BibitemOpen
  \bibfield  {author} {\bibinfo {author} {\bibfnamefont {J.}~\bibnamefont
  {Vukov}}, \bibinfo {author} {\bibfnamefont {A.}~\bibnamefont {Szolnoki}},\
  and\ \bibinfo {author} {\bibfnamefont {G.}~\bibnamefont {Szabó}},\ }\href
  {https://doi.org/10.1103/physreve.88.022123} {\bibfield  {journal} {\bibinfo
  {journal} {Physical Review E}\ }\textbf {\bibinfo {volume} {88}},\ \bibinfo
  {pages} {022123} (\bibinfo {year} {2013})}\BibitemShut {NoStop}%
\bibitem [{\citenamefont {Von~Neumann}(1966)}]{neumann1966theory}%
  \BibitemOpen
  \bibfield  {author} {\bibinfo {author} {\bibfnamefont {J.}~\bibnamefont
  {Von~Neumann}},\ }\href@noop {} {\emph {\bibinfo {title} {Theory of
  Self-Reproducing Automata}}}\ (\bibinfo  {publisher} {University of Illinois
  Press},\ \bibinfo {address} {USA},\ \bibinfo {year} {1966})\BibitemShut
  {NoStop}%
\bibitem [{\citenamefont {Ilachinski}(2001)}]{ilachinski2001cellular}%
  \BibitemOpen
  \bibfield  {author} {\bibinfo {author} {\bibfnamefont {A.}~\bibnamefont
  {Ilachinski}},\ }\href@noop {} {\emph {\bibinfo {title} {Cellular automata: a
  discrete universe}}}\ (\bibinfo  {publisher} {World Scientific Publishing
  Company},\ \bibinfo {year} {2001})\BibitemShut {NoStop}%
\bibitem [{\citenamefont {Ermentrout}\ and\ \citenamefont
  {Edelstein-Keshet}(1993)}]{ermentrout1993cellular}%
  \BibitemOpen
  \bibfield  {author} {\bibinfo {author} {\bibfnamefont {G.}~\bibnamefont
  {Ermentrout}}\ and\ \bibinfo {author} {\bibfnamefont {L.}~\bibnamefont
  {Edelstein-Keshet}},\ }\href {https://doi.org/10.1006/jtbi.1993.1007}
  {\bibfield  {journal} {\bibinfo  {journal} {Journal of Theoretical Biology}\
  }\textbf {\bibinfo {volume} {160}},\ \bibinfo {pages} {97–133} (\bibinfo
  {year} {1993})}\BibitemShut {NoStop}%
\bibitem [{\citenamefont {Nowak}\ and\ \citenamefont
  {May}(1992)}]{nowak1992evolutionary}%
  \BibitemOpen
  \bibfield  {author} {\bibinfo {author} {\bibfnamefont {M.~A.}\ \bibnamefont
  {Nowak}}\ and\ \bibinfo {author} {\bibfnamefont {R.~M.}\ \bibnamefont
  {May}},\ }\href {https://doi.org/10.1038/359826a0} {\bibfield  {journal}
  {\bibinfo  {journal} {Nature}\ }\textbf {\bibinfo {volume} {359}},\ \bibinfo
  {pages} {826–829} (\bibinfo {year} {1992})}\BibitemShut {NoStop}%
\bibitem [{\citenamefont {Alfaro}\ and\ \citenamefont
  {Sanjuán}(2024)}]{alfaro2024hamming}%
  \BibitemOpen
  \bibfield  {author} {\bibinfo {author} {\bibfnamefont {G.}~\bibnamefont
  {Alfaro}}\ and\ \bibinfo {author} {\bibfnamefont {M.~A.~F.}\ \bibnamefont
  {Sanjuán}},\ }\href {https://doi.org/10.1103/physreve.109.014203} {\bibfield
   {journal} {\bibinfo  {journal} {Physical Review E}\ }\textbf {\bibinfo
  {volume} {109}},\ \bibinfo {pages} {014203} (\bibinfo {year}
  {2024})}\BibitemShut {NoStop}%
\bibitem [{\citenamefont {Pikovsky}\ and\ \citenamefont
  {Politi}(2016)}]{pikovsky2016lyapunov}%
  \BibitemOpen
  \bibfield  {author} {\bibinfo {author} {\bibfnamefont {A.}~\bibnamefont
  {Pikovsky}}\ and\ \bibinfo {author} {\bibfnamefont {A.}~\bibnamefont
  {Politi}},\ }\href@noop {} {\emph {\bibinfo {title} {Lyapunov Exponents: A
  Tool to Explore Complex Dynamics}}}\ (\bibinfo  {publisher} {Cambridge
  University Press},\ \bibinfo {year} {2016})\BibitemShut {NoStop}%
\bibitem [{\citenamefont {Tsallis}\ \emph {et~al.}(1997)\citenamefont
  {Tsallis}, \citenamefont {Plastino},\ and\ \citenamefont
  {Zheng}}]{tsallis1997power}%
  \BibitemOpen
  \bibfield  {author} {\bibinfo {author} {\bibfnamefont {C.}~\bibnamefont
  {Tsallis}}, \bibinfo {author} {\bibfnamefont {A.}~\bibnamefont {Plastino}},\
  and\ \bibinfo {author} {\bibfnamefont {W.-M.}\ \bibnamefont {Zheng}},\ }\href
  {https://doi.org/10.1016/s0960-0779(96)00167-1} {\bibfield  {journal}
  {\bibinfo  {journal} {Chaos, Solitons \& Fractals}\ }\textbf {\bibinfo
  {volume} {8}},\ \bibinfo {pages} {885–891} (\bibinfo {year}
  {1997})}\BibitemShut {NoStop}%
\bibitem [{\citenamefont {Tirnakli}\ and\ \citenamefont
  {Borges}(2016)}]{tirnakli2016standard}%
  \BibitemOpen
  \bibfield  {author} {\bibinfo {author} {\bibfnamefont {U.}~\bibnamefont
  {Tirnakli}}\ and\ \bibinfo {author} {\bibfnamefont {E.~P.}\ \bibnamefont
  {Borges}},\ }\href {https://doi.org/10.1038/srep23644} {\bibfield  {journal}
  {\bibinfo  {journal} {Scientific Reports}\ }\textbf {\bibinfo {volume} {6}},\
  \bibinfo {pages} {23644} (\bibinfo {year} {2016})}\BibitemShut {NoStop}%
\bibitem [{gnu(2023)}]{gnuplot-doc}%
  \BibitemOpen
  \href@noop {} {\bibinfo {title} {{Gnuplot Documentation}}},\ \bibinfo
  {howpublished} {\url{http://www.gnuplot.info/documentation.html}} (\bibinfo
  {year} {2023}),\ \bibinfo {note} {accessed: 17 July 2023}\BibitemShut
  {NoStop}%
\bibitem [{\citenamefont {Bazeia}\ \emph
  {et~al.}(2021{\natexlab{a}})\citenamefont {Bazeia}, \citenamefont {Ferreira},
  \citenamefont {Oliveira},\ and\ \citenamefont
  {Szolnoki}}]{bazeia2021environment}%
  \BibitemOpen
  \bibfield  {author} {\bibinfo {author} {\bibfnamefont {D.}~\bibnamefont
  {Bazeia}}, \bibinfo {author} {\bibfnamefont {M.~J.~B.}\ \bibnamefont
  {Ferreira}}, \bibinfo {author} {\bibfnamefont {B.~F.~d.}\ \bibnamefont
  {Oliveira}},\ and\ \bibinfo {author} {\bibfnamefont {A.}~\bibnamefont
  {Szolnoki}},\ }\href {https://doi.org/10.1038/s41598-021-91994-7} {\bibfield
  {journal} {\bibinfo  {journal} {Scientific Reports}\ }\textbf {\bibinfo
  {volume} {11}},\ \bibinfo {pages} {1} (\bibinfo {year}
  {2021}{\natexlab{a}})}\BibitemShut {NoStop}%
\bibitem [{\citenamefont {Bazeia}\ \emph
  {et~al.}(2021{\natexlab{b}})\citenamefont {Bazeia}, \citenamefont
  {Bongestab}, \citenamefont {de~Oliveira},\ and\ \citenamefont
  {Szolnoki}}]{bazeia2021effects}%
  \BibitemOpen
  \bibfield  {author} {\bibinfo {author} {\bibfnamefont {D.}~\bibnamefont
  {Bazeia}}, \bibinfo {author} {\bibfnamefont {M.}~\bibnamefont {Bongestab}},
  \bibinfo {author} {\bibfnamefont {B.}~\bibnamefont {de~Oliveira}},\ and\
  \bibinfo {author} {\bibfnamefont {A.}~\bibnamefont {Szolnoki}},\ }\href
  {https://doi.org/10.1016/j.chaos.2021.111255} {\bibfield  {journal} {\bibinfo
   {journal} {Chaos, Solitons \& Fractals}\ }\textbf {\bibinfo {volume}
  {151}},\ \bibinfo {pages} {111255} (\bibinfo {year}
  {2021}{\natexlab{b}})}\BibitemShut {NoStop}%
}
\end{thebibliography}
\end{document}